# *In Situ* Melting and Revitrification as an Approach to Microsecond Time-Resolved Cryo-Electron Microscopy

Jonathan M. Voss, Oliver F. Harder, Pavel K. Olshin, Marcel Drabbels, and Ulrich J. Lorenz*

**Affiliation:** Laboratory of Molecular Nanodynamics, École Polytechnique Fédérale de Lausanne, 1015 Lausanne, Switzerland

* To whom correspondence should be addressed. E-mail: ulrich.lorenz@epfl.ch

**Abstract**

Proteins typically undergo conformational dynamics on the microsecond to millisecond timescale as they perform their function, which is much faster than the time-resolution of cryo-electron microscopy and has thus prevented real-time observations. Here, we propose a novel approach for microsecond time-resolved cryo-electron microscopy that involves rapidly melting a cryo specimen *in situ* with a laser beam. The sample remains liquid for the duration of the laser pulse, offering a tunable time window in which the dynamics of embedded particles can be induced in their native liquid environment. After the laser pulse, the sample vitrifies in just a few microseconds, trapping particles in their transient configurations, so that they can subsequently be characterized with conventional cryo-electron microscopy. We demonstrate that our melting and revitrification approach is viable and affords microsecond time resolution. As a proof of principle, we study the disassembly of particles after they incur structural damage and trap them in partially unraveled configurations.

Proteins perform a range of tasks that are quintessential for life. They are involved in harvesting energy, maintaining homeostasis, catalyzing metabolic reactions, reproduction, and sensing stimuli. As nanoscale machines, proteins are dynamic systems whose free energy surface determines their motions and characteristic timescales[1]. Backbone motions and side-chain rotations involve transitions between minima separated by small barriers and typically occur on picosecond to nanosecond timescales, leading to fast, small-amplitude fluctuations of the protein structure[2]. In contrast, the dynamics associated with the activity of a protein frequently involve large-amplitude motions of entire domains that occur on a timescale of microseconds to milliseconds[1,2].

Much of our knowledge about protein function derives from static structures at atomic resolution, which frequently allow one to infer function, particularly if structures of different functional states can be obtained[3–6]. However, in the absence of real-time observations, our understanding of proteins is necessarily incomplete. Directly observing proteins as they perform their tasks, in real time and at atomic resolution, holds the promise of fundamentally advancing our grasp of these nanoscale machines[1].

X-ray crystallography has long been the most important tool in structural biology[7]. Recently, the advent of bright x-ray pulses has also made it possible to carry out time-resolved experiments, even with ultrafast time resolution[8,9]. However, the requirement of a crystalline sample poses a specific challenge for studying the dynamics of proteins, as the crystal packing hinders large-amplitude motions[10–12]. In contrast, single-particle cryo-electron microscopy (cryo-EM)[13–19] enables the observation of biological specimens in their frozen hydrated state[20,21]. The revolution that the field has recently undergone has also created new opportunities for obtaining time-resolved information[22,23]. In time-resolved cryo-EM, dynamics are induced by mixing two reactants[24–26] or exposing the sample to a short light pulse[27,28]. Short-lived intermediates are then trapped in vitreous ice by promptly plunge freezing the sample. However, with a time resolution of several milliseconds[22], which is ultimately limited by the timescales for plunging (~5 ms) (Ref. 28) and vitrification (~1 ms) (Ref. 21), this technique is currently too slow to capture many relevant processes.

Here, we propose a novel approach for carrying out time-resolved cryo-EM with microsecond time resolution. The concept (Figure 1a–e) is based on rapidly melting a vitrified sample *in situ* with a heating laser (Figure 1a). Once in their native liquid environment at room temperature (Figure 1b), the embedded particles are subjected to an external stimulus that induces dynamics (Figure 1c). For example, a second laser pulse can be used to photoactivate a caged compound and liberate ATP or a small peptide that interacts with the protein[29,30]. In the simplest case, the heating laser itself can be used to create a temperature jump to initiate conformational dynamics. At a well-defined point in time, the heating laser is switched off to induce rapid revitrification and trap the particles in their transient configurations (Figure 1d), so that they can be subsequently characterized with established cryo-EM techniques (Figure 1e).

It is not obvious that implementing such a scheme should be possible. Usually, warming up a cryo sample is scrupulously avoided in order to prevent devitrification. Here, we demonstrate that this obstacle can be overcome. We employ cryo samples with a holey gold film[31] supported by a gold mesh. For some experiments, we additionally sandwich the sample between two sheets of few layer graphene, as illustrated in Figure 1f (Note S1 and Figure S1), so as to reduce evaporation of the liquid water in the vacuum of the microscope[32]. The sample is heated *in situ* with a laser pulse of tens of microseconds duration (532 nm, 24 μm FWHM spot size), which we obtain by chopping the output of a continuous laser with an acousto-optic modulator[33] (Note S3 and Figure S2). The gold film and graphene sheets strongly absorb at the laser wavelength and serve as the heat source to melt the ice. In contrast, neither the proteins nor the ice absorb the laser radiation, which prevents photodamage to the specimen. In all experiments, we have centered the heating laser onto the area under observation.

**Results and Discussion**

**Rapid *in situ* melting and vitrification**

We first demonstrate that it is indeed possible to melt a cryo sample *in situ* and subsequently achieve rapid vitrification. To this end, we intentionally prepare crystalline ice samples, which allows us to directly verify the success of the experiment. Figure 2a shows a thin ice sheet held at 100 K that is suspended over a hole of the gold film and enclosed between graphene layers. Bend contours are visible, indicating that the ice sheet is crystalline. This is confirmed by the diffraction pattern, which exhibits the signature of hexagonal ice[21] (Figure 2b). When we irradiate the sample with an individual 20 µs laser pulse with a power of 25 mW, the diffraction pattern remains largely unchanged (Figure 2c), indicating that the sample has not melted. Laser pulses of up to 63 mW yield the same result (Figure 2d–g). However, a single pulse with a power of 75 mW provides sufficient heat to induce melting (Figure 2h). Most of the ice sheet now exhibits homogeneous contrast, with the diffraction pattern revealing the signature of vitreous ice[21] (Figure 2i, selected area marked by the green circle in Figure 2h). In the top left of the sample, a small void has formed. We observe such voids if defects in the graphene sheets allow liquid water to evaporate. The lighter contrast on the right side of the sample indicates an area where some ice has remained crystalline, as a selected area diffraction pattern reveals (Figure 2i inset, selected area indicated with a grey circle in Figure 2h). Evidently, at this laser power, the ice sheet barely reaches the melting point and remains partially crystalline. Incrementally increasing the laser power thus provides an *in situ* calibration of the power required for melting.

For the experiment illustrated in Figure 3a–d, the threshold laser power for melting and vitrification was first calibrated in a different area of the sample. When we use this power to heat the crystalline sample in Figure 3a,b, melting and vitrification is achieved with a single laser pulse (Figure 3c,d). The uniform contrast suggests that here, the ice film suspended across the hole has completely vitrified. We note that the laser power needed to induce melting is lower than in Figure 2, where the sample is in close proximity to the bars of the TEM grid. Heat transfer simulations agree with this observation, showing that the required power increases closer to the bars (Note S4 and Figure S3).

**Characterization of the temperature evolution and time resolution**

For the purpose of time-resolved cryo-EM experiments, it is crucial to have a detailed understanding of the temperature evolution of the sample, which ultimately determines the attainable time resolution. We therefore characterize the heating and cooling dynamics *in situ* with time-resolved electron microscopy[34–38] (Note S5 and Figure S2). We heat the sample in Figure 4a, which is held at 100 K, with a 20 µs laser pulse (13 mW). To probe its temperature at a specific point in time, we record a time-resolved diffraction pattern of a selected area of the gold film with a 1 ns electron pulse (~$10^4$ electrons). The diffraction intensity of the gold film decreases exponentially with temperature (Debye-Waller behavior[39]) and therefore provides a suitable probe of the sample temperature. We then repeat the experiment stroboscopically (1 kHz repetition rate) to probe the temperature at different points in time. Since a thin layer of ice would not be stable under illumination with thousands of laser pulses, we perform the experiment in the area shown in Figure 4a, which is not covered by ice. Nevertheless, the temperature evolution of the bare gold film provides a good approximation of the heating and cooling dynamics of a cryo sample. As discussed in detail below, simulations confirm that the presence of a thin ice film barely alters the temperature evolution. Moreover, the ice and the underlying gold film have near-identical temperatures.

Figure 4b shows a diffraction pattern of the holey gold film collected from the area marked with a red dot in Figure 4a, onto which we have also aligned the heating laser. As the laser warms the sample, the diffraction intensity decreases quickly, as shown for the (331), (420), and (422) reflections in Figure 4c. After only a few microseconds, the temperature in the selected area stabilizes, increasing only slightly over time. When the heating laser is switched off at 20 µs, the sample promptly recools as it dissipates heat towards its surroundings, which have remained at cryogenic temperature. From exponential fits (solid red curve, Note S6,7), we obtain heating and cooling times of 1.2 µs and 3.6 µs, respectively. Similar timescales are obtained in different areas of the sample (orange and yellow dots in Figure 4a and Figure S4e,f).

Heat transfer simulations (Note S4 and Figure S3) reproduce these timescales reasonably well, yielding 1.0 µs heating and cooling times (Figure 4d, solid red curve). If we add a thin layer of vitreous ice, its temperature (blue curves) closely follows that of the underlying gold film, with which it is in close contact. With increasing ice thickness, the heating and cooling times become longer. However, the temperature at which the sample stabilizes during the laser pulse barely changes. We find that the sample temperature plateaus at 300 K for a laser power of 35 mW (Figure 4d), which is in reasonable agreement with the experimentally determined power that induces melting.

It may appear counterintuitive that melting a cryo sample *in situ* should result in vitrification, since heating up a vitreous sample ordinarily causes irreversible crystallization[40]. The characterization of the temperature evolution reveals that this is made possible by the high heating and cooling rates of nearly $10^8$ K/s, which is more than two orders of magnitude faster than what is required to outrun crystallization and achieve vitrification[41,42]. Several factors enable such fast cooling. The tightly focused laser beam heats the sample only locally, so that after the laser pulse, the heat can be rapidly dissipated to adjacent areas that have remained at cryogenic temperature. Our simulations also reveal that the bars of the specimen grid play an important role in this respect. Due to their large heat capacity, they barely warm up and therefore act as an effective heat sink. Crucially, the cooling rate also determines how fast the dynamics of embedded particles can be arrested, which occurs when the sample is cooled below the glass transition temperature of water at 137 K (Refs. [43,44]). We conclude that for ice thicknesses suitable for cryo-EM, our approach yields a time resolution of 5 µs or better, three orders of magnitude faster than what is currently possible with time-resolved cryo-EM[22]. Importantly, this matches the time resolution of cryo-EM to the characteristic timescale on which the relevant dynamics of most proteins occur[1,2]. Snapshots of the dynamics at different points in time can then be acquired by simply changing the duration of the heating laser pulse.

**Proof of principle experiment**

As a proof of principle, we study the rapid disassembly of the GroEL protein complex[45] in response to structural damage that we induce through electron beam irradiation. It is well-known that proteins are heavily damaged during cryo-imaging with typical electron doses[46,47]. However, the cryogenic matrix traps fragments in place and thus limits structural degradation, which would otherwise render high-resolution imaging impossible[46,47]. In a liquid environment, however, structural changes can freely occur. Melting a cryo sample *in situ* after it has been exposed to a significant electron dose should therefore allow the damaged proteins to unravel.

Figure 5a shows a cryo sample of GroEL on a holey gold film, imaged with a dose of 14 electrons/Å$^2$. Here, we have illuminated only the top part, so that particles in the unexposed area do not incur any beam damage and can serve as a control. For simplicity, the sample is not enclosed between graphene sheets. Even though this allows some water to evaporate during laser heating, it is still possible to carry out experiments with laser pulse durations of tens of microseconds without evaporating the entire sample. In order to allow dynamics of the beam-damaged GroEL to occur, we melt the sample with a 10 µs laser pulse, after which it revitrifies, arresting the particles in their transient configurations (Figure 5b). The irradiated area (dashed line) appears lighter than the remainder of the sample, a phenomenon that we consistently observe, but that is absent without electron beam exposure. This suggests that this area has thinned as volatile radiolysis products trapped in the ice[47] were liberated during the melting process (see also Figure S5). The proteins are no longer visible in this area, indicating that they have disintegrated, leaving only fragments that offer little contrast. A few larger fragments are visible near the edge of the irradiated area (red arrows). The proteins in the bottom part of the image, which had not been exposed to the electron beam, have remained intact. This demonstrates that the disintegration of GroEL particles we observe is caused by electron beam irradiation and not the melting and revitrification process.

The effect of exposure to the electron beam appears to extend beyond the irradiated area. In its immediate vicinity, the particle density is lower, and partially disassembled proteins as well as fragments are visible (red arrows). A possible explanation is that melting liberates highly reactive radiolysis products trapped in the ice, such as OH and H radicals, protons, and solvated electrons[47]. The diffusion lengths of these species within the laser pulse duration of 10 µs are hundreds of nanometers[48], suggesting that they will reach intact GroEL particles and damage them. It is also possible that convective flow of the water film occurs as areas of different thickness equilibrate. With increasing duration of the heating laser pulse, convection seems to play a larger role since the spatial distributions of intact particles and fragments are increasingly reshuffled, as shown for a 30 µs laser pulse in Figure S5. After irradiation with a 50 µs laser pulse (Figure 5c,d), the ice thickness is almost homogeneous across the hole in the gold film, with intact particles and fragments visible in both the exposed and unexposed areas (Figure 5d, insets).

We note that the disintegration of the proteins observed in Figure 5d does not result from the laser irradiation, but is only present if the sample has first been exposed to the electron beam. This is evidenced by Figure S6, which shows a neighboring hole of the gold film that had not been irradiated by the electron beam. The particles in this hole are intact, even though they were exposed to a near-identical laser intensity and reached a similar temperature (the FWHM of the Gaussian laser spot, which is centered on the sample, is more than an order of magnitude larger than the distance separating the holes).

Lowering the electron dose reduces the damage inflicted on the GroEL particles, so that their disintegration proceeds more slowly, allowing us to take snapshots of different stages. If we irradiate the sample with only 1.4 electrons/Å$^2$ (Figure S7), the particles do not disintegrate entirely when the sample is melted with a 10 µs laser pulse (Figure 5e). Instead, a large number of damaged proteins and fragments remain visible in the exposed area. Increasing the laser pulse duration to 20 µs allows the disassembly to proceed further (Figure 5f). Few particles remain that resemble intact GroEL, while the size of the fragments has further decreased. We note that we have adjusted the laser power with the procedure described in Figure 2, so that we can be certain that the plateau temperature of the sample exceeds the melting point of ice. Moreover, heat transfer simulations show that for the particular sample geometry used here, evaporative cooling limits the plateau temperature, so that it should not exceed room temperature for the laser powers employed (Figure S8).

**Conclusion**

We have demonstrated that it is possible to rapidly laser melt and revitrify a cryo sample *in situ*, so that particle dynamics can occur in liquid for a well-defined amount of time. This opens up the perspective of carrying out microsecond time-resolved cryo-EM experiments. In a proof of principle experiment, we have observed the disassembly of GroEL following electron beam damage, which reveals a surprising insight about cryo-imaging. Even at a dose as low as 1.4 electrons/Å$^2$, damage to the particles is so extensive that once the vitreous ice matrix that preserves their shape is melted, they completely disintegrate. These initial results suggest that our method may be applicable to a wider range of processes. For instance, it is straightforward to implement temperature jump experiments by adjusting the laser power to heat the sample well above room temperature, which should make it possible to study folding and unfolding dynamics. Another possibility is to trigger dynamics with an additional laser pulse once the sample reaches room temperature. In the case of a photoactive protein[8,49], the laser pulse itself can initiate the dynamics.

Alternatively, it can provide a biomimetic trigger by dissociating a photo-activated caged compound to release ions, ATP, or small peptides, or to induce a pH jump[29,30,50]. Such experiments could be simplified by activating the photorelease compounds already in the vitreous state of the sample. Since the proteins are unable to undergo dynamics in the vitreous ice matrix, the trigger can only become active once the sample is melted. We have in fact demonstrated such a scheme in the experiments of Figure 5, where the trigger, electron beam irradiation, was applied before *in situ* melting. Precise control of the sample temperature will be crucial for such experiments to succeed. The laser power needed to reach a given sample temperature can be calibrated *in situ* by determining the threshold power for melting (Figure 2). Fortunately, small variations in the thickness of the ice layer do not change the plateau temperature of the sample (Figure 4d). This is particularly true for samples not enclosed in graphene, for which evaporative cooling stabilizes the temperature.

...3
**Associated Content**

**Supporting Information:** Notes for sample preparation, *in situ* melting and vitrification, heat transfer simulations, characterization of the temperature evolution and time resolution with time-resolved EM, analysis of the time-resolved diffraction patterns, and fitting procedure for the heating and cooling times; figures for sample preparation for graphene-enclosed cryo samples, sketch of modified transmission electron microscope, heat transfer simulations, time-resolved diffraction data, time-resolved cryo-EM GroEL experiments.

**Author Information**

**Corresponding author:** Ulrich J. Lorenz

**E-mail:** ulrich.lorenz@epfl.ch

**Notes:** The authors declare no competing financial interests.

**Acknowledgements**

The authors acknowledge Christoph Schillai for his input on sample preparation and for critically reading the manuscript. We kindly thank the Centre Interdisciplinaire de Microscopie Electronique and the Plateforme Technologique de la Production et Structure des Protéines at EPFL for use of equipment. We thank D. Demurtas and K. Lau for their support and helpful discussions. This work was supported by the ERC Starting Grant 759145 and by the Swiss National Science Foundation Grant PP00P2_163681.

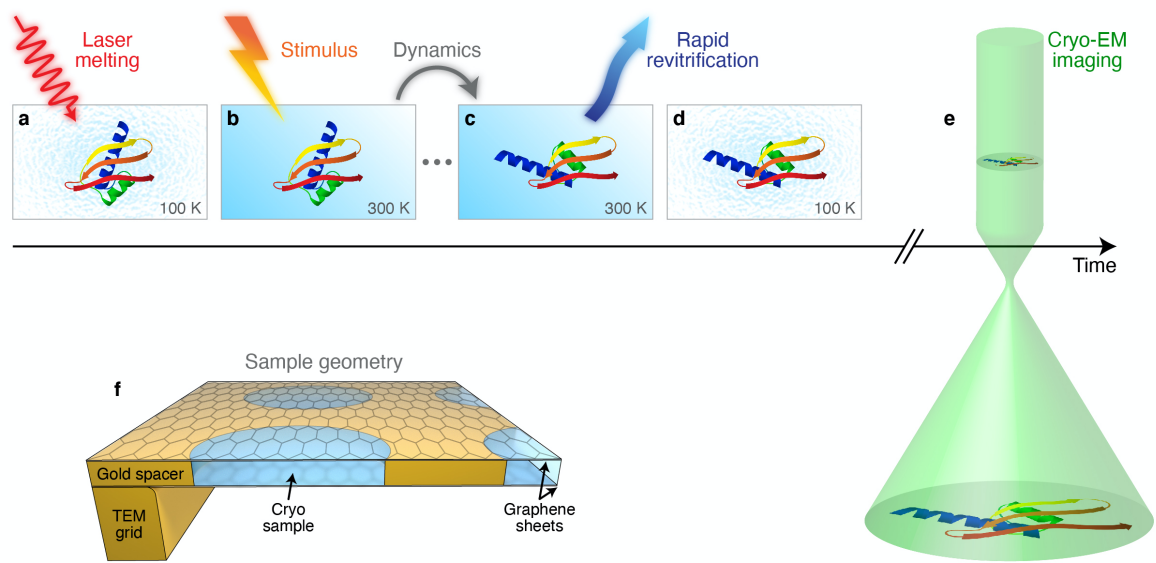

**Figure 1. Experimental concept and sample geometry.** (a–e) Illustration of rapid *in situ* melting and revitrification for time-resolved cryo-EM. (a) A cryo sample is melted *in situ* with a heating laser. (b) Once the sample is liquid and has reached room temperature, dynamics of the embedded particles are induced with an external stimulus, *e.g.* a short second laser pulse that releases a caged compound such as ATP or a peptide. (c) While the particle undergoes conformational changes, the heating laser is switched off at a given point in time so that the sample rapidly cools and revitrifies. (d) The particle is trapped in its transient configuration and can be subsequently imaged with conventional cryo-EM techniques (e). (f) Illustration of the sample geometry. A cryo sample supported by a holey gold film is enclosed between two graphene layers, which prevent evaporation in the vacuum of the microscope as the sample is melted *in situ* with a laser pulse.

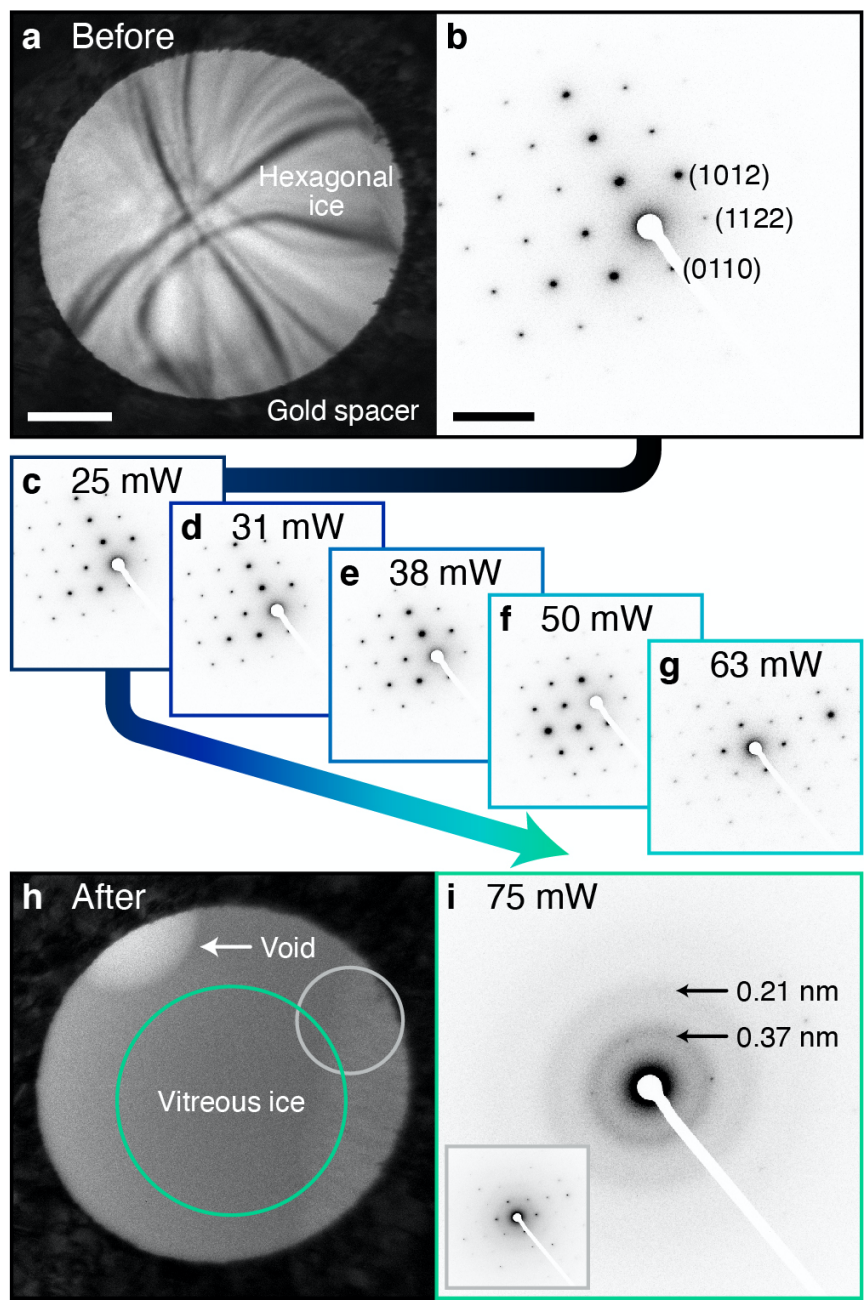

**Figure 2**. **Determination of the laser power required for *in situ* melting and vitrification.** (a) Micrograph and (b) selected area diffraction pattern of a thin film of hexagonal ice. (c–g) The sample is irradiated with individual laser pulses of 20 μs duration. As the laser power is increased up to 63 mW, the diffraction pattern remains largely unchanged, indicating that melting has not occurred. (h,i) A single pulse with a power of 75 mW provides sufficient heat to melt the sample, which then rapidly vitrifies after the end of the laser pulse. Scale bars, 300 nm and 4 nm$^{-1}$.

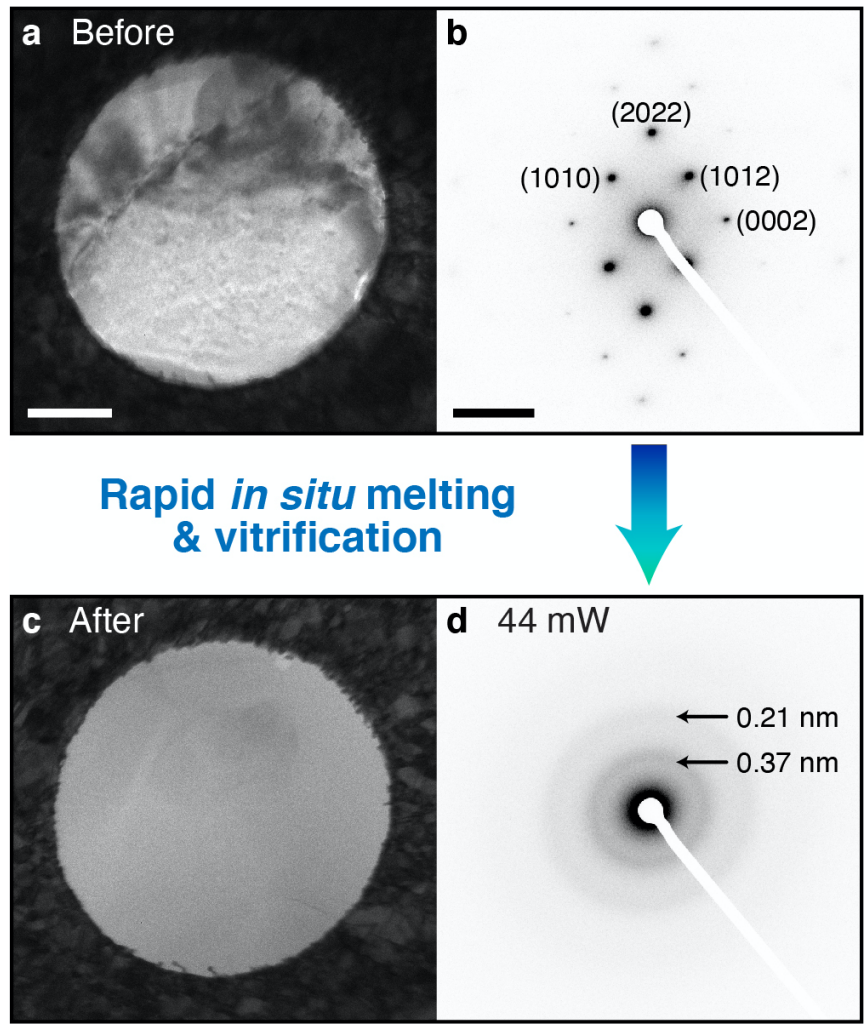

**Figure 3. *In situ* melting and vitrification with a single laser pulse.** (a) Micrograph and (b) selected area diffraction pattern of a thin film of hexagonal ice. (c,d) After calibration of the laser power in a different area, the sample is successfully melted and vitrified with a single laser pulse. Scale bars, 300 nm and 4 nm$^{-1}$.

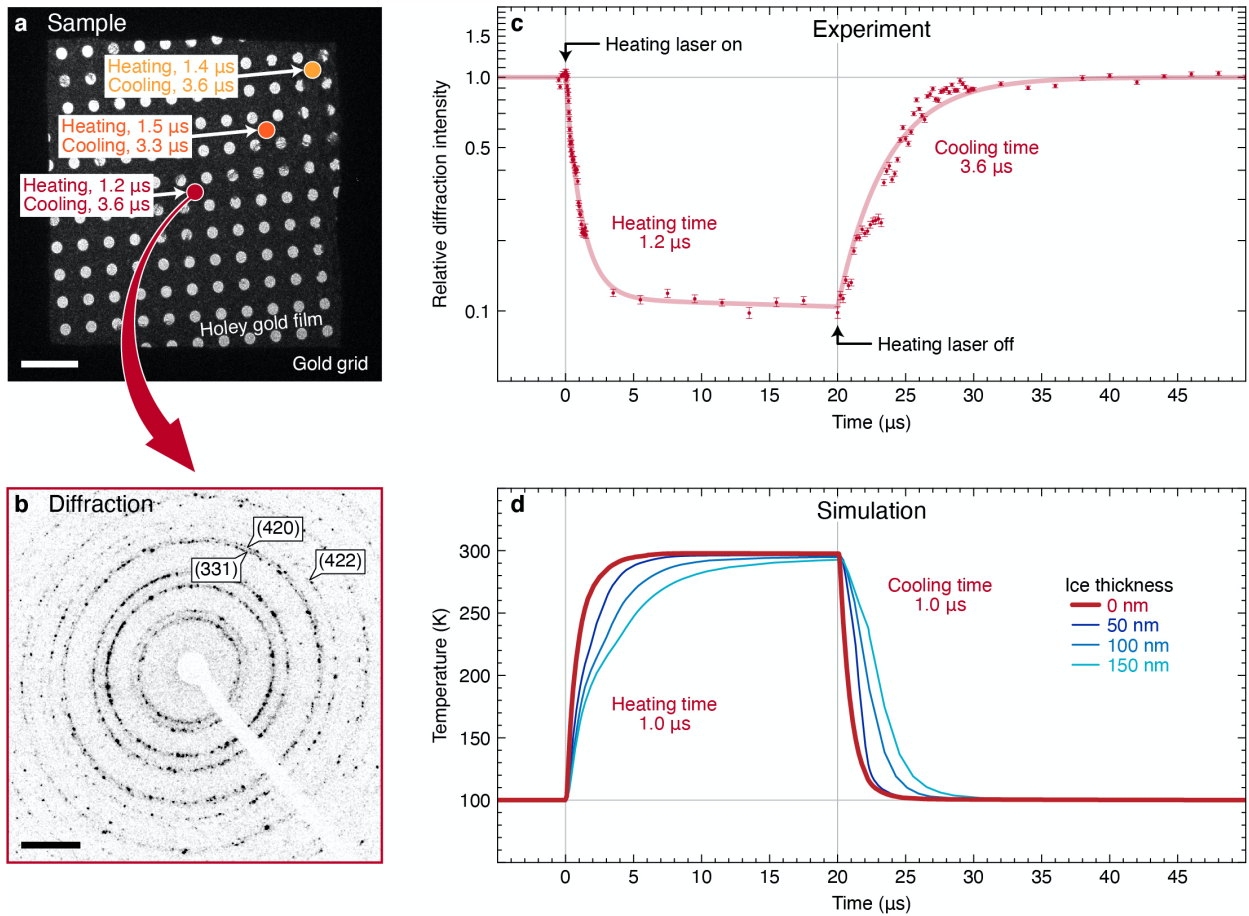

**Figure 4. Characterization of the temperature evolution of the sample under illumination with microsecond laser pulses.** (a) Micrograph of the sample in an area that is free of ice. Heating and cooling timescales of the gold film were measured at three different positions (dots), onto which the laser was focused. Scale bar, 5 µm. (b) Selected area diffraction pattern of the area marked with a red dot in (a). Scale bar, 5 nm$^{-1}$. (c) Temporal evolution of the intensity of the (331), (420), and (422) reflections in (b) under irradiation with a 20 µs laser pulse (dots). Heating and cooling times of 1.2 µs and 3.6 µs, respectively, are determined from fits with exponential functions (solid curve). The error bars represent the standard error. (d) A simulation of the temperature evolution of the sample in (c) yields similar heating and cooling times of 1.0 µs (solid red curve). In the presence of a layer of vitreous ice (blue curves), heating and cooling times increase, while the maximum temperature at which the sample stabilizes barely changes.

## Disassembly dynamics of GroEL (high electron dose)

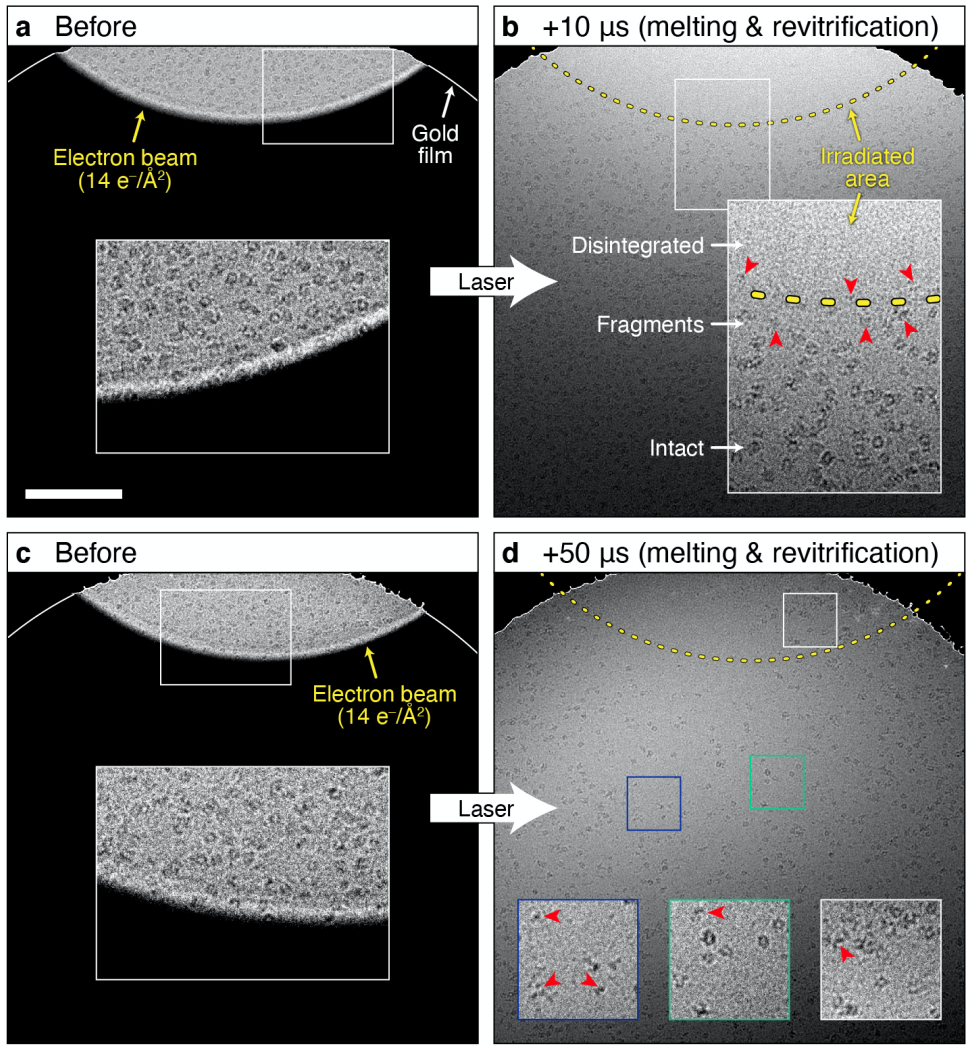

## Disassembly dynamics of GroEL (low electron dose)

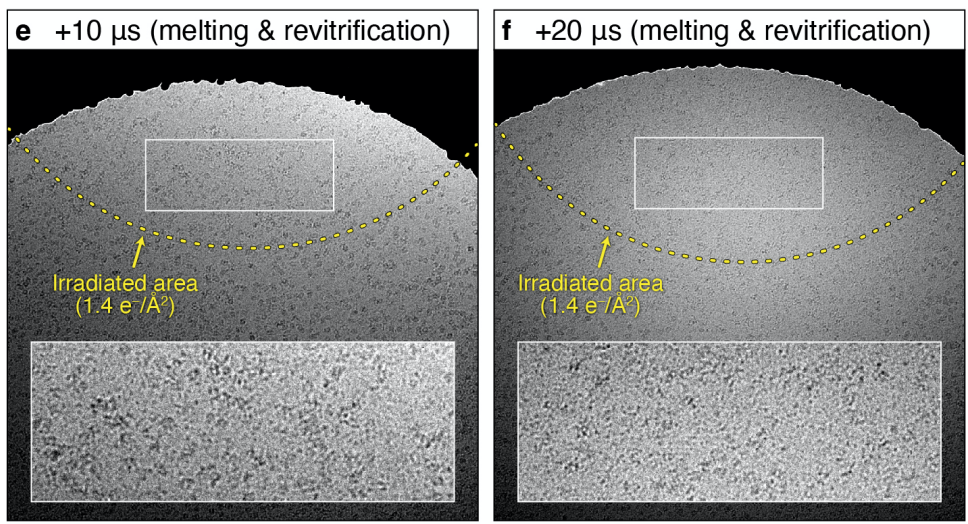

**Figure 5. Microsecond time-resolved cryo-EM of the rapid disassembly of GroEL following electron beam damage.** (a) Micrograph of GroEL on a holey gold film. Only the top part is illuminated with the electron beam (14 electrons/Å$^2$) in order to limit beam damage to particles in that area. The lines sketch the outline of the hole of the gold film. (b) The sample is melted with a 10 µs laser pulse *in situ*, allowing the damaged GroEL particles to unravel. After the laser pulse, the sample revitrifies, trapping particles in their transient configurations. The dashed line delineates the sample area that was irradiated in (a). The red arrows highlight protein fragments that are visible at the boundary of the irradiated area. (c,d) The experiment in (a,b) is repeated with a 50 µs laser pulse. (e,f) The electron beam damage inflicted on the GroEL particles is reduced by lowering the electron dose to 1.4 electrons/Å$^2$. Melting the sample with a laser pulse of 10 µs (e) and 20 µs duration (f) traps the particles in different stages of the disassembly process. In the insets, the contrast has been adjusted to make the particles more easily visible. Scale bar, 250 nm.

# Supporting Information for

# *In Situ* Melting and Revitrification as an Approach to Microsecond Time-Resolved Cryo-Electron Microscopy

Jonathan M. Voss, Oliver F. Harder, Pavel K. Olshin, Marcel Drabbels, and Ulrich J. Lorenz*

**Affiliation:** Laboratory of Molecular Nanodynamics, École Polytechnique Fédérale de Lausanne, 1015 Lausanne, Switzerland

* To whom correspondence should be addressed. E-mail: ulrich.lorenz@epfl.ch

**This PDF file includes:**

Notes S1–7

Figures S1–8

References

**Note S1. Graphene-enclosed cryo sample preparation**

The graphene-enclosed cryo samples are prepared with the procedure illustrated in Figure S1. The gold spacer is obtained by evaporating 30 nm of gold onto a Quantifoil TEM grid (R1.2/1.3 holey carbon film on 200 mesh copper, Figure S1a,b). The coated grid is placed onto the surface of a concentrated aqueous solution of $(NH_4)_2S_2O_8$ for 15 minutes to etch away the copper mesh and is subsequently transferred to a bath of deionized water for 10 minutes to remove the etchant (Figure S1c). The carbon/gold film is then transferred onto bilayer CVD graphene on copper foil (ACS Material, Figure S1d). To remove water and improve the contact between the graphene and amorphous carbon layers, the assembly is gently heated on a hotplate for 10 minutes at 50 °C. Once dried, the entire assembly is floated on the surface of a $(NH_4)_2S_2O_8$ etching solution for approximately 2 hours to dissolve the copper foil (Figure S1e). Once the copper has been completely removed, the assembly is transferred to a deionized water bath for 10 minutes in order to remove the etchant. Next, a 600 mesh gold TEM grid is submerged into the water and delicately drawn out to suspend the assembly onto the TEM grid (Figure S1f). The grid is then dried on a hotplate for 10 minutes at 50 °C, after which a 10 µL drop of deionized water is placed on the surface of the gold film. Another multilayer graphene film (3-5 layers, ACS Material) is floated on the surface of a deionized water bath. The TEM grid is submerged in the water and withdrawn to suspend the graphene across the gold film and thus trap water in the holes of the gold film (Figure S1g). The grid assembly is then blotted with filter paper for approximately 10 seconds to remove excess water and is immediately plunge frozen (Figure S1h) and loaded into a cryo specimen holder.

**Note S2. Preparation of cryo samples of GroEL**

Cryo samples of GroEL were prepared on Quantifoil UltraAuFoil TEM grids (R1.2/1.3 holey gold film on 300 mesh gold or R2/2 holey gold film on 200 mesh gold, 50 nm film thickness). The grids were plasma cleaned for 30 seconds using an ELMO glow discharge system operating with negative glow discharge head polarity, 0.860 mA plasma current, and 0.2 mBar residual air pressure. A 3 µL drop of a 1.3 mg/mL GroEL solution (Takara Bio Europe SAS, used without further purification) was placed on each of the plasma cleaned grids. The grids were inserted into a Vitrobot MarkIV (Thermo Fisher Scientific) held at 100% relative humidity and 22 °C and were blotted with 595 filter paper for 2.5 seconds with a blotting force of -15. Immediately after blotting, the samples were vitrified by plunge freezing in liquid ethane.

**Note S3. Rapid *in situ* melting and vitrification**

All experiments were carried out with a modified JEOL 2200FS transmission electron microscope operating at 160 kV accelerating voltage[1] (Figure S2). To induce rapid melting and vitrification, the sample is irradiated *in situ* with microsecond laser pulses (532 nm, tens of microseconds) that are obtained by chopping the output of a continuous laser with an acousto-optic modulator (20–80% rise and fall times of 73 ns and 134 ns, respectively, see Figure S3c for the pulse shape). The laser beam is directed at the sample by means of a mirror located above the upper pole piece of the objective lens and strikes it at close to normal incidence. A 25 cm lens focuses the laser beam to a spot size of 24±1 μm FWHM, as measured by a knife edge scan at the sample location.

**Note S4. Heat transfer simulations**

Finite element heat transfer simulations of the temperature evolution of the sample under irradiation with microsecond laser pulses were performed with COMSOL Multiphysics. The simulation geometry is shown in Figure S3a,b. A gold film (30 nm thickness) rests on top of an amorphous carbon substrate (20 nm thickness), which is supported by a 600 mesh gold TEM grid (14.5 μm wide and 10 μm thick bars, 30 μm x 30 μm viewing area). The amorphous carbon/gold film features a regular pattern of holes (1.2 μm diameter, 2.8 μm pitch) and supports a thin layer of vitreous ice (0–150 nm, Figure S3b). In order to reduce computational cost, we have omitted the graphene sheets that enclose the cryo samples in our experiments. As the thickness of the graphene layers are small, their heat capacity and thermal conductivity are negligible compared to the remainder of the assembly[2], so that omitting them in the simulation does not change the temperature evolution. We have, however, considered that the graphene layers absorb the heating laser radiation and have included this contribution in our calculation of the total absorption cross section of the sample. The simulation area is limited to a square of 122.5 μm side length, which is more than five times larger than the laser spot size (24 μm FWHM, indicated by empty circles in Figure S3a). At the boundaries of the simulated area, we fix the temperature to the initial temperature of the geometry (100 K) in order to account for the large heat capacity of the remainder of the sample. We note that even if we do not apply this boundary condition, the temperature evolution is virtually identical.

We use literature values for the temperature dependent heat capacity and thermal conductivity of gold[3] as well as for the thermal conductivity of amorphous carbon[4]. Since reliable low-temperature values for the heat capacity of amorphous carbon are not available[5], we use its room temperature value[6] and scale it to low temperatures assuming the temperature dependence for graphite[7]. We note that when we change the heat capacity of amorphous carbon in a wide range, the temperature evolution changes only nominally. The heat capacity and thermal conductivity of supercooled water are unknown in a wide temperature range known as 'no man's land', where rapid crystallization has so far prevented the study of its thermal properties[8,9]. We therefore use the heat capacity of supercooled water in silica nanopores instead, which is available for the whole temperature range[10]. For the thermal conductivity of water, we use its room temperature value, which is also close to that of amorphous ice at 100 K (Ref. 11).

The temperature of the entire geometry is initially set to 100 K. Heating with a 20 µs laser pulse is then simulated by placing a Gaussian heat source (24 µm FWHM) on the top surface of the gold film, with the heating rate calculated from the incident laser power and the absorption of both gold[12] and graphene[13,14]. For the temporal structure of the 20 µs heating laser pulse, we use the experimentally determined pulse shape (black curve in Figure S3c). In Figure 4D and Figure S3c, the heating laser power (35 mW) is chosen such that the temperature at the center of the laser focus plateaus at ~300 K when the laser beam is centered on the simulation geometry. We simulate three different positions of the laser focus (solid dots in Figure S3a) and probe the evolution of the sample temperature in their center (Figure S3c, no added layer of ice). While heating and cooling times remain virtually unchanged (~1 µs), the plateau temperature of the sample decreases as the laser focus approaches the bars of the TEM grid. Therefore, the distance of the sample to the bars of the TEM grid has to be taken into account when the laser power required for melting is calibrated *in situ*, as shown in Figure 2. In the simulations reported in Figure 4d, the laser beam was centered on the simulation geometry (red dot in Figure S3a). For simulations that include a layer of vitreous ice, we report the temperature of the ice, not the temperature of the amorphous carbon/gold film, although they are nearly identical.

In order to understand how evaporative cooling affects the temperature evolution of the sample during irradiation with microsecond laser pulses, we perform heat transfer simulations using the experimental geometry from Figure 5. A gold film (50 nm thickness) is supported by a 200 mesh gold TEM grid (38.5 μm wide and 15 μm thick bars, 86.8 μm x 86.8 μm viewing area). The gold film features a regular pattern of holes (2 μm diameter, 4 μm pitch). A thin layer of vitreous ice (100 nm thickness) fills the holes and covers both sides of the gold film. The simulation area is limited to a square of 190 μm side length. To account for the large heat capacity of the specimen grid, the gold bars extend 435 μm past the simulation area in each direction. The temperature of the entire geometry is initially set to 100 K.

Heating with a 15 μs laser pulse is simulated by placing a Gaussian heat source (24 μm FWHM) on the top surface of the gold film in the center of the simulation geometry. To account for evaporative cooling, we apply a negative heat source to both surfaces of the water film. The cooling rate is determined from the temperature-dependent enthalpy of evaporation[15] and the temperature-dependent evaporation rate[16] (Figure S8a), which we calculate using literature values of the vapor pressure[17] assuming an evaporation coefficient of $\gamma$ = 1 (Ref 9).

We probe the temperature evolution of vitreous ice located in the middle of the central hole in the gold film. Figure S8b shows the heating and cooling dynamics of the ice film for different laser powers. With a 40 mW laser pulse, the ice just barely surpasses the melting point. As the laser power is increased further, the plateau temperature of the sample increases only slightly, reaching 284 K at 120 mW. Evaporative cooling acts as a negative feedback that leads to a small variation in plateau temperature for a large range of laser powers.

**Note S5. Characterization of the temperature evolution and time resolution with time-resolved EM**

The temperature evolution of the sample under irradiation with microsecond laser pulses (Figure 4) was characterized with time-resolved electron microscopy. This technique uses short electron pulses to capture processes that are faster than the time resolution of the electron camera[18]. For these experiments, the temperature of the emitter is lowered until electron emission ceases. Short electron pulses (1 ns, ~$10^4$ electrons) are then generated by irradiating the Schottky emitter of the microscope with a UV laser pulse (266 nm, 1 ns, 200 nJ pulse energy). A precisely timed electron pulse is then used to record a selected area diffraction pattern of the sample at a specific point in time during irradiation with a microsecond laser pulse. The experiment is repeated stroboscopically in order to acquire a range of time frames and thus capture the entire heating and cooling dynamics of the sample. Every stroboscopic diffraction pattern was acquired with 8,000 electron pulses, and the dynamics at each position of the heating laser were recorded at least five times.

For the stroboscopic diffraction experiments, we reduced the power of the heating laser to 13 mW in order to minimize small irreversible deformations of the sample that slowly occur under exposure to thousands of laser pulses and that interfere with the stroboscopic experiment (see below). Our heat transfer simulations confirm that while different laser powers result in different plateau temperatures of the sample, they barely change the heating and cooling timescales, which are the figures of merit that we extract from the experiment.

**Note S6. Analysis of the time-resolved diffraction patterns**

For our analysis of the temperature evolution of the sample, we monitor the intensity of the (331), (420), and (422) reflections of the gold film (Figure 4b). We note that if we choose different reflections, we obtain very similar, albeit noisier transients of the diffraction intensity.

The intensity of each selected diffraction spot is determined by integrating a rectangular region of interest centered around the diffraction spot (green boxes in Figure S4c) and subtracting the intensity of the diffraction background, as determined from the average intensity of the areas marked in blue boxes in

Figure S4c. Over the course of a stroboscopic experiment (several hours), the electron beam intensity slowly varies and causes the intensities of the diffraction spots to change accordingly. To account for this effect, the diffraction intensities are normalized by the electron beam intensity, which we estimate with the following procedure. We calculate the total intensity of the diffraction spots included in our analysis for time frames recorded before time zero and after 40 µs, when the sample temperature has returned to 100 K. We then use a spline of this total diffraction intensity to estimate the intensity the electron beam had when each individual time-resolved diffraction pattern was recorded. After normalization, we average the intensity of each diffraction spot over all delay scans and calculate the corresponding standard error. Both the average and error are weighted by the relative electron beam intensity of the frame. Finally, the average intensities of the diffraction spots are summed up to give the transients shown in Figure 4c and Figure S4d–f. The reported errors are obtained through error propagation.

In our analysis, we have excluded diffraction spots whose intensity changes irreversibly over the course of the stroboscopic experiment or that yield a very poor signal-to-noise ratio. We note that when this exclusion is not made, the transients obtained are very similar, albeit noisier, and the heating and cooling timescales we extract are virtually identical. Irreversible changes of the intensity of some diffraction spots occur as the gold film slowly deforms under exposure to thousands of laser pulses, which causes some grains to slowly move into or out of the diffraction condition. As noted above, we have reduced the power of the heating laser to minimize such effects. Diffraction spots that nevertheless exhibit such irreversible changes are excluded from the analysis, as they would otherwise distort the stroboscopic measurement.

We have also excluded diffraction spots that exhibit obvious non-Debye-Waller behavior and whose intensity increases as the sample is being heated, instead of decreasing. Such behavior is frequently observed in nanocrystalline thin films and occurs as individual grains move into or out of the diffraction condition under the strain that laser heating induces. These effects may still be present for the diffraction spots included in our analysis. For a large enough sample, such effects should cancel out, so that Debye-Waller behavior is observed on average. However, this is not the case for the small number of grains in our experiment that fulfill the diffraction condition within the selected area (1.7 µm diameter). It is therefore important to note that one cannot easily draw conclusions about the magnitude of the induced temperature jump from the observed drop in diffraction intensity. For the same reason, the relative intensity drop observed in the measurements of Figure S4d–f does not accurately reflect that the plateau temperature is highest in the center of the gold film in Figure S4a. However, the determination of the heating and cooling times remains unaffected. We also note that if we do not exclude the diffraction spots whose intensity increases with temperature, the heating and cooling timescales do not change significantly.

**Note S7. Fitting procedure for the heating and cooling times**

In order to determine the heating and cooling times from the time-resolved diffraction experiments, we fit the logarithm of the relative diffraction intensity, $\log(I/I_0)$, which the Debye-Waller effect predicts to decrease linearly with temperature[19]. Here, $I$ is the diffraction intensity and $I_0$ is the diffraction intensity before time zero. We fit the heating dynamics ($\leq$ 20 μs) with the empirical piecewise function $f_1(t)$,

$$\log\left(\frac{I}{I_0}\right) = f_1(t) = \begin{cases} 0 & \text{for } t \leq 0 \\ -a(1 - e^{-t/\tau_{heating}}) - bt & \text{for } t > 0 \text{ and } t \leq 20 \text{ μs} \end{cases}$$

where $\tau_{heating}$ is the heating time, and $a$ and $b$ are fit parameters. For the cooling dynamics after the end of the laser pulse at $t = 20$ μs, we fit the transient with the function $f_2(t)$,

$$\log\left(\frac{I}{I_0}\right) = f_2(t) = f_1(t = 20 \text{ μs}) \cdot e^{-(t - 20 \text{ μs})/\tau_{cooling}} \qquad \text{for } t \geq 20 \text{ μs}$$

where $\tau_{cooling}$ is the cooling time.

In the stroboscopic experiments, we sampled the rapid heating and cooling of the sample with shorter time steps in order to better capture these fast dynamics. Without accounting for the uneven spacing of the data points, a least-squares fit would be biased towards the regions of high sampling rate. We therefore employ weighting factors in the fit that are inversely proportional to the standard error of the data point and inversely proportional to the sampling rate. The sampling rate is determined from the time steps between each data point and its nearest neighbors.

The heating and cooling times in the heat transfer simulations are determined in an analogous manner by fitting the temperature change, $T(t) - T(t < 0)$, with the functions $f_1(t)$ and $f_2(t)$ as described above.

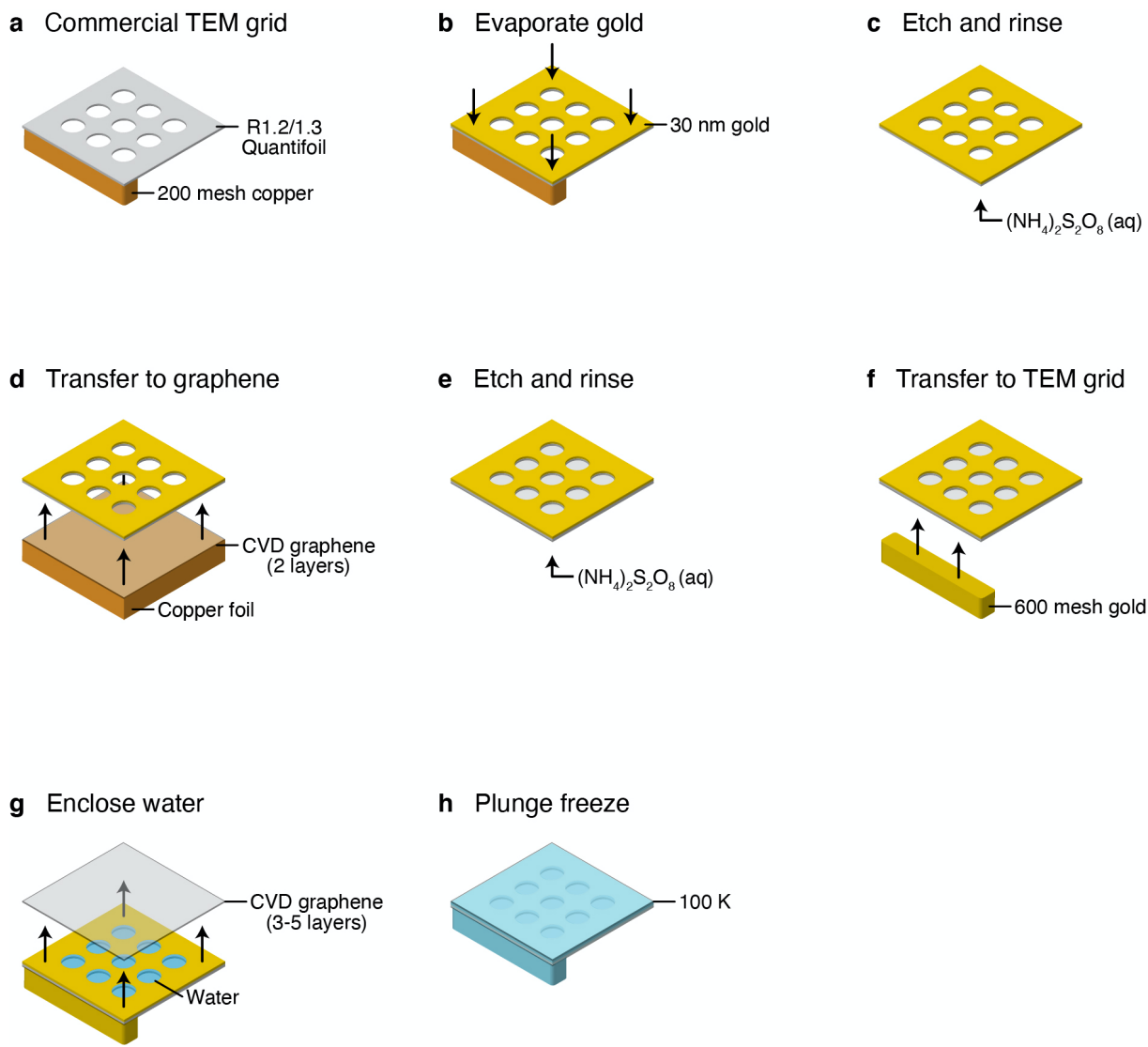

**Figure S1. Sample preparation workflow for graphene-enclosed cryo samples.** (a,b) A Quantifoil TEM grid is coated with gold. (c) The copper mesh of the TEM grid is etched away and the gold film is rinsed. (d) The film is placed onto bilayer graphene on copper foil. (e) The copper foil is etched away and the film is rinsed. (f) The film is transferred onto a gold TEM grid and (g) water is enclosed with graphene. (h) The assembly is plunge frozen.

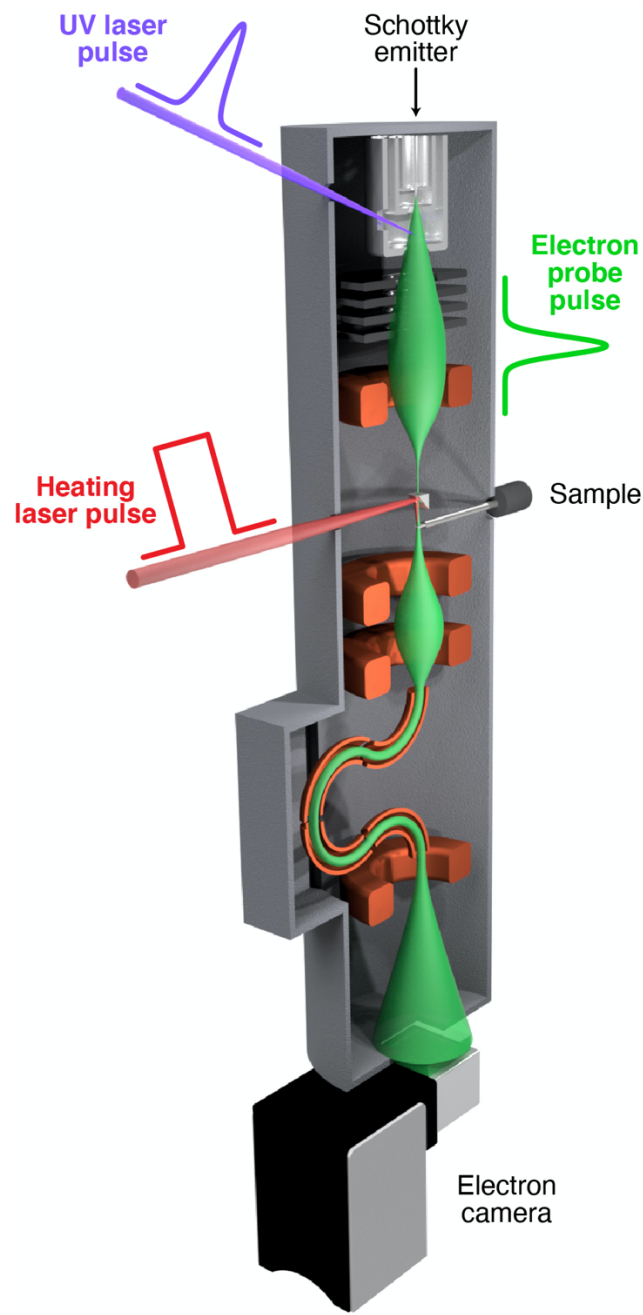

**Figure S2. Sketch of the modified transmission electron microscope.** The sample is heated *in situ* with a laser pulse of tens of microseconds duration, obtained by chopping the output of a continuous laser with an acousto-optic modulator. In the stroboscopic experiments characterizing the temperature evolution of the sample (Figure 4 of the main text and Figure S4), electron probe pulses are generated by illuminating the Schottky emitter of the microscope with a UV laser pulse.

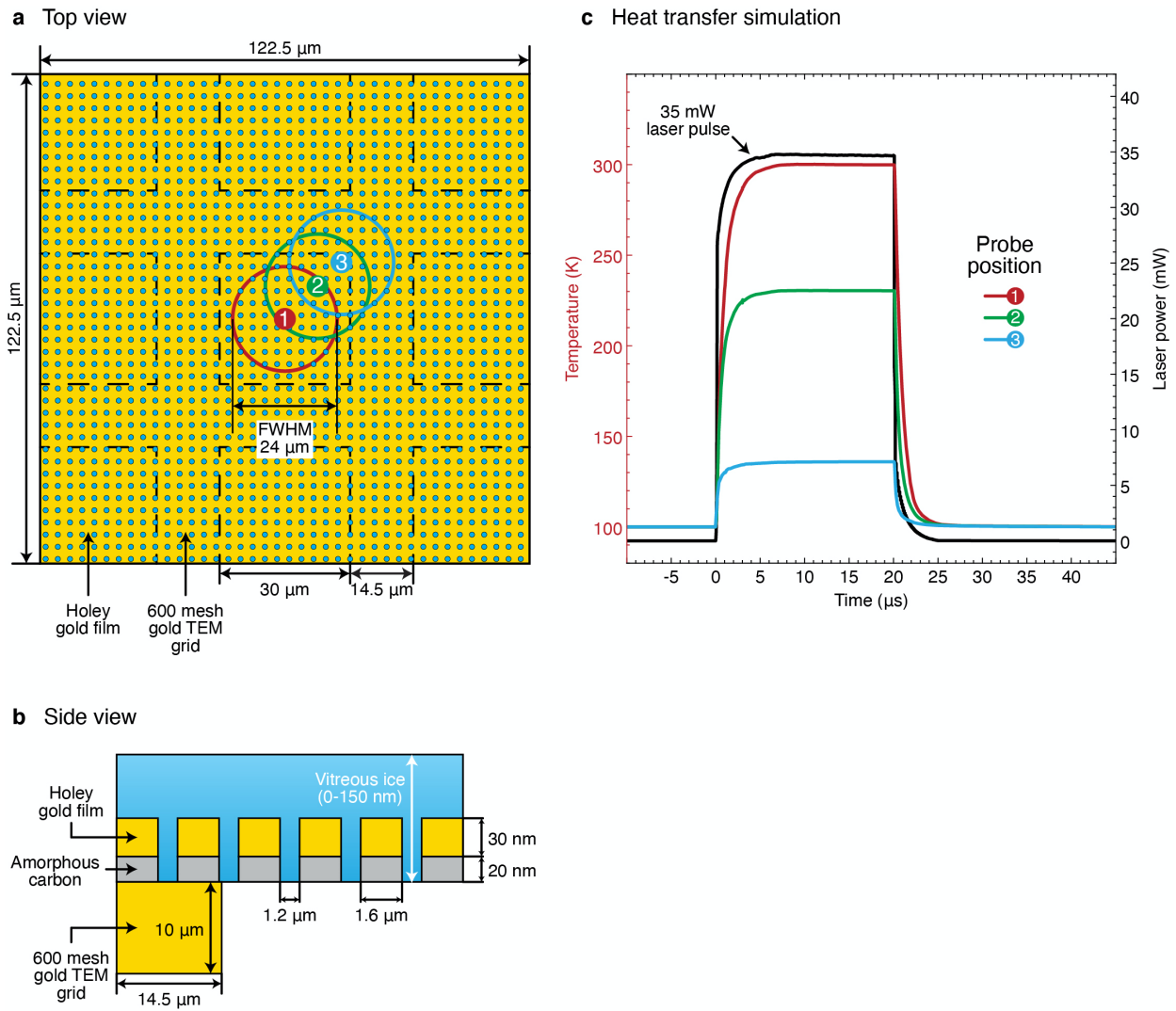

**Figure S3. Heat transfer simulations.** (a,b) Top and side view of the simulated geometry, respectively. The empty circles mark the position of the laser focus, and the dots mark the regions probed in the corresponding simulation. (c) Temporal profile of the laser pulse (black curve) and temperature evolution of the gold film. While the magnitude of the temperature jump depends on the distance to the grid bars, the heating and cooling times (~1 μs) are largely unchanged.

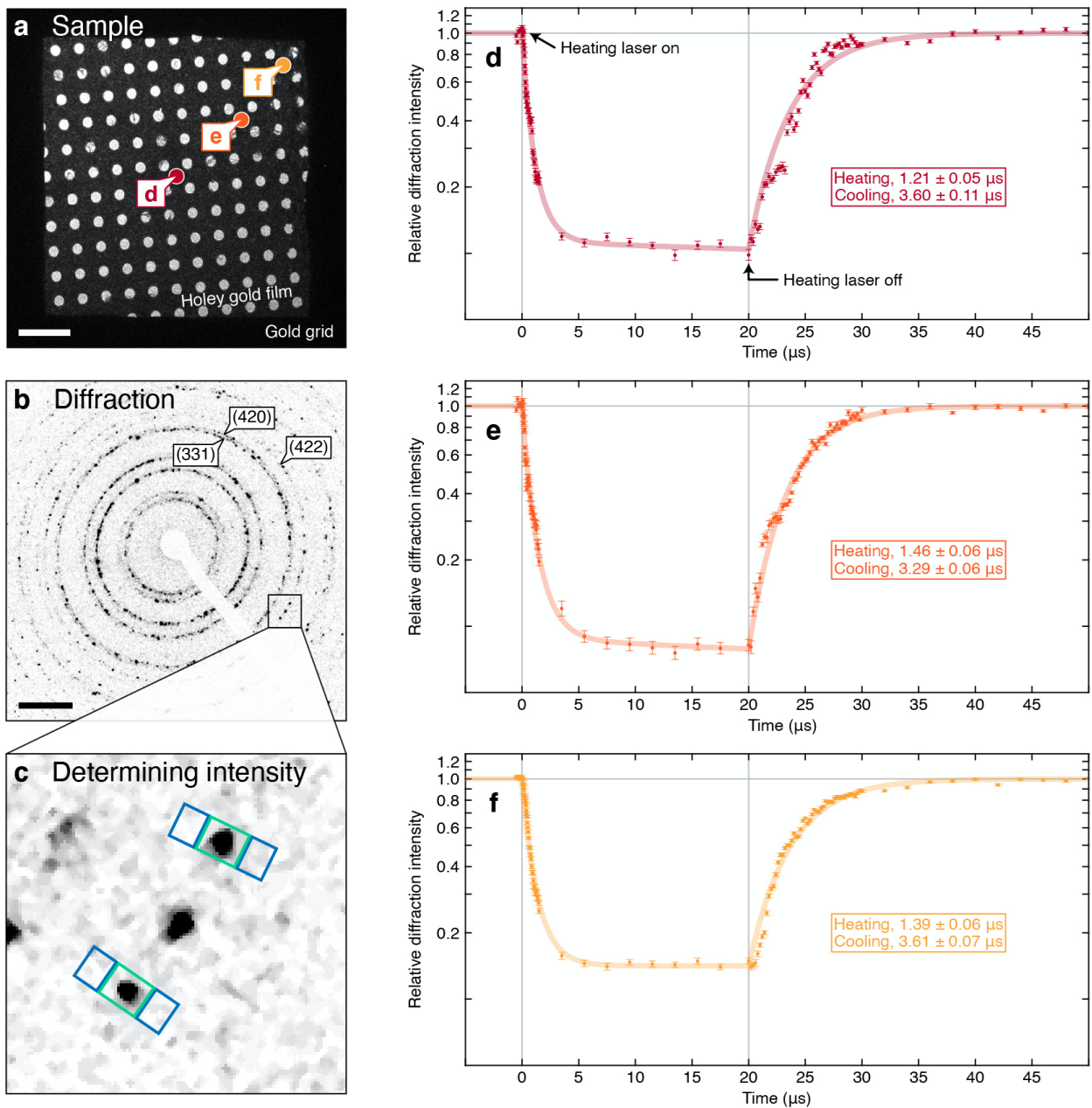

**Figure S4. Characterization of the temperature evolution of the sample under illumination with microsecond laser pulses.** (a) Micrograph of the sample in a region that is free of ice. The dots mark the selected areas from which diffraction patterns were acquired and onto which the laser was focused. Scale bar, 5 µm. (b) Selected area diffraction pattern of the area marked with a red dot in (a). Scale bar, 5 nm$^{-1}$. (c) The intensities of diffraction spots are determined by integrating a rectangular region of interest around the diffraction spot (green boxes) and subtracting the intensity of the diffraction background, as determined

from the average intensity of the areas marked with blue boxes. (d–f) Temporal evolution of the diffraction intensity at the sample positions marked in (a). Heating and cooling times indicated in the figure are determined from fits with exponential functions (solid lines, see Methods). The error bars represent the standard error.

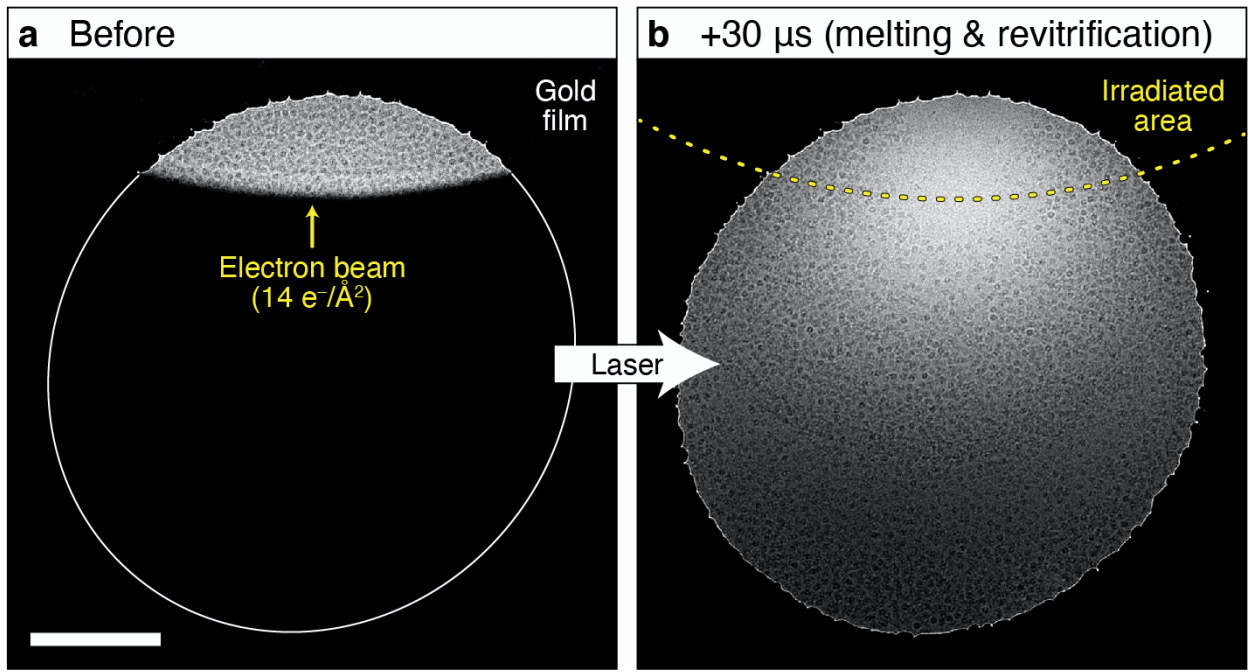

**Figure S5. Microsecond time-resolved cryo-EM of the rapid disassembly of GroEL following electron beam damage.** (a) Micrograph of a vitrified sample of GroEL on a holey gold film. Only the top portion of the sample is irradiated with the electron beam (14 electrons/Å$^2$) in order to limit beam damage to particles in that area. The line sketches the outline of the hole of the gold film. (b) The sample is melted with a 30 μs laser pulse *in situ*, allowing the damaged GroEL particles to unravel. Following the laser pulse, the sample revitrifies, trapping particles in their transient configurations. The dashed line marks the sample area that was irradiated in (a). Unlike in Figure 5b of the main text, intact particles can be seen in the irradiated area. Moreover, the particle density is depleted beyond the region that was damaged by the electron beam. This suggests that convection has redistributed the particles. Scale bar, 300 nm.

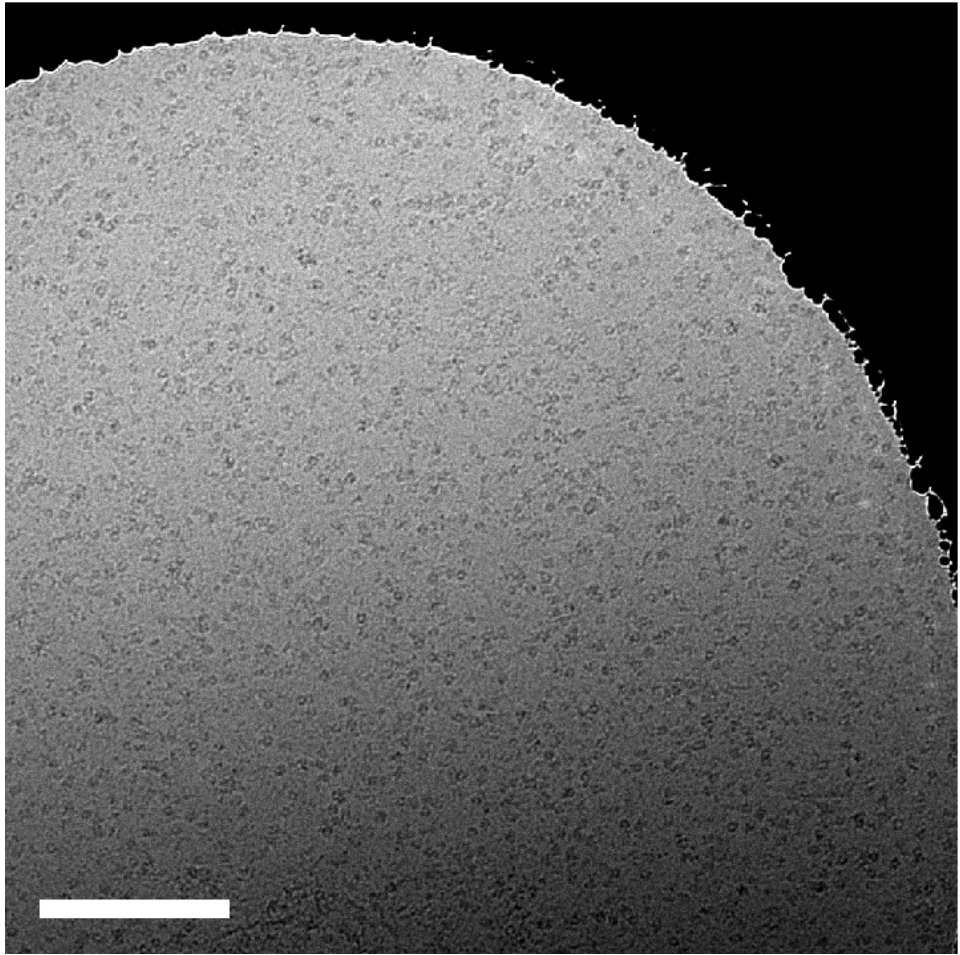

**Figure S6. GroEL particles in the hole adjacent to the area shown in Figure 5c,d.** The micrograph shows GroEL particles in a neighboring hole, which had not been irradiated by the electron beam in Figure 5c. Despite being exposed to a near-identical laser intensity and reaching a similar temperature, the particles in this hole are still intact, indicating that the dynamics observed in Figure 5d are not the result of laser irradiation. Scale bar, 250 nm.

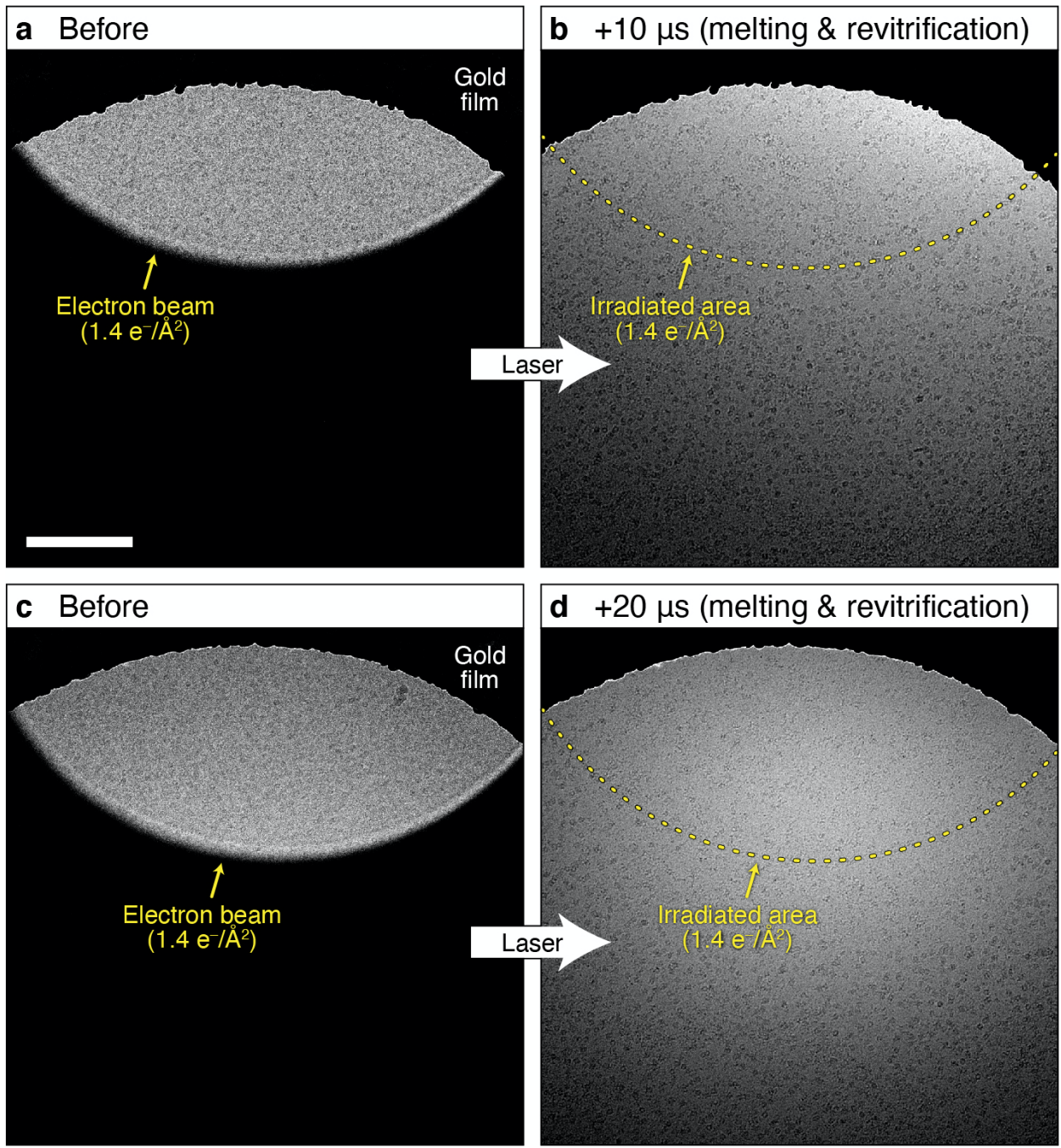

**Figure S7. Microsecond time-resolved cryo-EM of the rapid disassembly of GroEL following electron beam damage.** (a) Micrograph of a vitrified sample of GroEL on a holey gold film. Only the top portion of the sample is irradiated with the electron beam (1.4 electrons/Å²) in order to limit beam damage to particles in that area. (b) The sample is melted with a 10 µs laser pulse *in situ*, allowing the damaged GroEL particles to unravel. Following the laser pulse, the sample revitrifies, trapping particles in their transient configurations. The dashed line marks the sample area that was irradiated in (a). (c,d) The experiment in (a,b) is repeated with a 20 µs laser pulse. Scale bar, 250 nm.

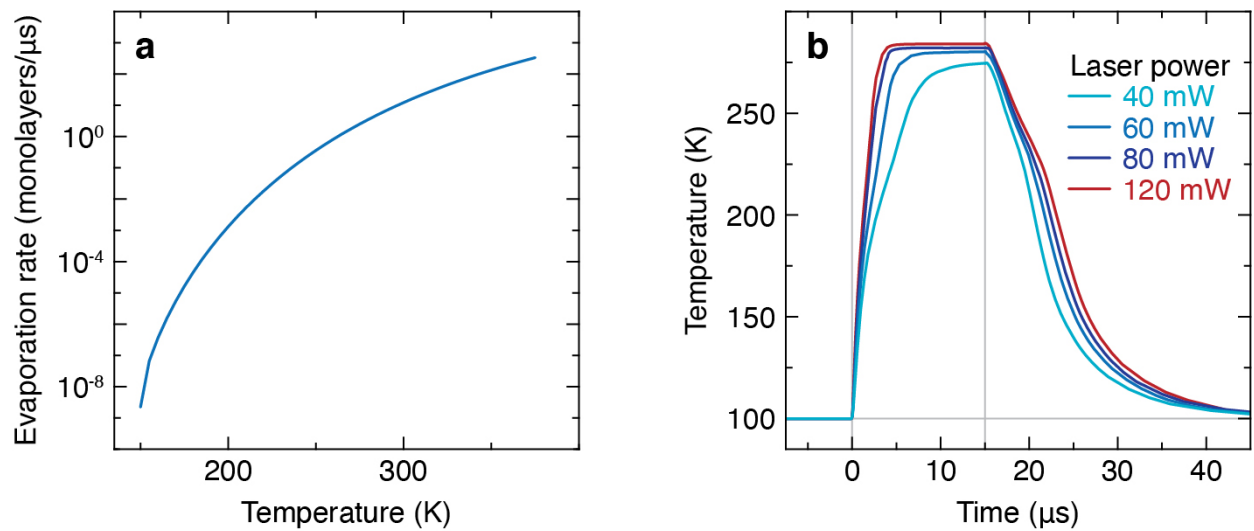

**Figure S8. Simulations of the temperature evolution in the experiments in Figure 5.** (a) Temperature-dependent evaporation rate of water (from Ref. 17). (b) Simulated temperature evolution of an ice film heated by a 15 µs laser pulse of different laser powers with evaporative cooling of the sample included.